\pdfoutput=1

\documentclass[3p,times,twocolumn]{elsarticle}

\usepackage{ecrc}

\volume{00}

\firstpage{1}

\journalname{Nuclear and Particle Physics Proceedings}

\runauth{Monika Kofarago for the ALICE collaboration}

\jid{nppp}

\jnltitlelogo{Nuclear and Particle Physics Proceedings}

\usepackage{amssymb}

 \usepackage{lineno}
 \usepackage{subfig}
 \usepackage[permil]{overpic}
 \usepackage{xspace}
 \usepackage{amsmath}
 \usepackage{multirow}
 \usepackage{booktabs}

\usepackage[figuresright]{rotating}

\newcommand{\pT}           {$p_{\,\rm T}$\xspace}
\newcommand{\Dphi}         {$\Delta\varphi$\xspace} 
\newcommand{\Deta}         {$\Delta\eta$\xspace} 
\newcommand{\pTtrig}       {$p_{\,\rm T,trig}$\xspace}
\newcommand{\pTassoc}      {$p_{\,\rm T,assoc}$\xspace}
\newcommand{\GeVc}         {~GeV/\textit{c}\xspace}
\newcommand{\snn}          {\mbox{$\sqrt{s_{\rm NN}}=2.76$~TeV}\xspace}

\begin{document}

\begin{frontmatter}


 \author{Monika Kofarago\fnref{label1,label2}}
 \ead{monika.kofarago@cern.ch}
 \fntext[label1]{for the ALICE collaboration}
 \fntext[label2]{CERN and MTA Wigner RCP}

\dochead{}

\title{Near-side jet peak broadening in Pb--Pb collisions at $\sqrt{s_{\rm NN}} = 2.76$ TeV}


\begin{abstract}
Two-particle angular correlation measurements are sensitive probes of the interactions of particles with the medium formed in heavy-ion collisions. Such measurements are done by determining the distribution of the relative pseudorapidity ($\Delta\eta$) and azimuthal angle ($\Delta\varphi$) of particles with respect to a higher $p_{\rm T}$ trigger particle ($1 < p_{\rm T,trig} < 8$~GeV/\textit{c}). The near-side peak is fitted with a function, which includes both the near-side jet peak and also accounts for the $\Delta\eta$-independent long-range correlations. The centrality evolution of the width (variance) of the fitted distribution is investigated. In Pb--Pb collisions a significant broadening of the near-side peak in the $\Delta\eta$ direction is observed from peripheral to central collisions, while in the $\Delta\varphi$ direction the peak is almost independent of centrality. For the 10\% most central events, a departure from the Gaussian shape is also observed at low transverse momentum ($1<p_{\rm T,assoc}<2$~GeV/\textit{c}, $1<p_{\rm T,trig}<3$~GeV/\textit{c}). In this contribution the results obtained by the ALICE experiment in Pb--Pb and pp collisions at \snn are shown, and they are interpreted in terms of radial and elliptic flow by comparing them to AMPT model simulations.
\end{abstract}

\begin{keyword}
Angular correlations \sep two-particle correlations \sep near-side peak \sep Pb--Pb \sep 2.76 TeV \sep ALICE


\end{keyword}

\end{frontmatter}

\section{Motivation}
In high-energy collisions, the scattered high \pT partons produce parton showers, which hadronize into collimated jets of hadrons. As the partons propagate through the expanding medium, they lose energy by medium induced gluon radiation and elastic scatterings, which is commonly referred to as jet quenching. This can result in the energy imbalance of back to back jets, with the radiated energy reappearing at large angles from the jet axis~\cite{Aad:2010bu, Chatrchyan:2011sx}. 

At low \pT, jets are difficult to identify in the large fluctuating background, and instead statistical analysis techniques are used to study the interaction of the propagating partons with the medium. One such technique is the measurement of two-particle angular correlations, where the distribution of the azimuthal angle difference (\Dphi) and pseudorapidity difference (\Deta) is studied on a statistical basis. In these measurements jets manifest themselves as a peak around $(\Delta\varphi,\Delta\eta) = (0,0)$ on the near-side, while as an approximately \Deta independent structure on the away-side (around \Dphi = $\pi$). The interaction of jets with the medium can modify the shape of the near-side jet peak~\cite{Armesto:2004pt, Armesto:2004vz, Romatschke:2006bb}, and such a centrality dependent broadening was seen by the STAR collaboration in Au--Au collisions at $\sqrt{s_{\rm NN}}=200$~GeV~\cite{Agakishiev:2011st}. In the present contribution, an extension of these measurements to LHC energies by the ALICE detector is discussed. The full details of the presented analysis can be found in \cite{Adam:2016ckp,Adam:2016tsv}.

\section{Analysis}
In the current analysis Pb--Pb and pp data at \snn recorded by the ALICE detector~\cite{Aamodt:2008zz} are studied. The Inner Tracking System (ITS) and the Time Projection Chamber (TPC) are used for tracking, while the ITS is also used for triggering purposes together with the V0 detector. The data is divided into five centrality classes according to the signal amplitude in the V0~\cite{Abelev:2013qoq}. More details about the event and track selection can be found elsewhere~\cite{Adam:2016ckp,Adam:2016tsv}.

In two-particle angular correlation measurements, all particles in a certain \pTtrig range are taken as trigger particles and are paired with all particles from a certain \pTassoc range. The selected ranges can be different or the same, in which case \pTtrig $>$ \pTassoc. From the \Dphi and \Deta difference of the pairs, the per trigger yield histograms are filled:
\begin{equation}
  \frac{1}{N_{trig}}\frac{d^2N_{assoc}}{d\Delta\varphi d\Delta\eta} = \frac{S(\Delta\varphi,\Delta\eta)}{M(\Delta\varphi,\Delta\eta)}
\end{equation}
The signal histogram is calculated from the particle pairs after an efficiency correction, which has a negligible effect on the shape of the distributions. The detector acceptance effects and pair inefficiencies are corrected by dividing the signal distribution by the so-called mixed event distribution. This is constructed in the same way as the signal distribution, but the associated particle is taken from a different event than the trigger particle, which destroys all physical correlations between them. The mixed event is normalized to unity at $(\Delta\varphi,\Delta\eta) = (0,0)$, which accounts for the fact that two particles traveling in the same direction have the same detection probability. The per trigger yield histograms are calculated in bins of the z-position of the collision vertex to account for the changing $\eta$ acceptance, and are afterwards added with the number of triggers as weights. 

Two particles traveling in approximately the same direction can be incorrectly reconstructed as one particle only, which results in a lower reconstruction efficiency around $(\Delta\varphi,\Delta\eta) = (0,0)$. This effect is corrected for by rejecting tracks, which have $|\Delta\eta|<0.02$ and \mbox{$|\Delta\varphi^*|<0.02$}, where $\Delta\varphi^*$ is the minimum azimuthal distance of the two particles after correcting for the bending in the magnetic field. This cut is applied to both the same and the mixed event histograms.

Correlated particles originating from the decay of a neutral particle are removed from the analysis by a cut in the invariant mass around the mass of $K_S^0$ and the $\Lambda$ particles. Particles originating from $\gamma$ conversions are also removed by an invariant mass cut. The exact details of the cuts can be found in~\cite{Adam:2016ckp,Adam:2016tsv}.

\begin{figure}[!htbp]
  \makebox[0.5\textwidth][c]{
    \subfloat[]{%
    \begin{overpic}[width=0.25\textwidth]{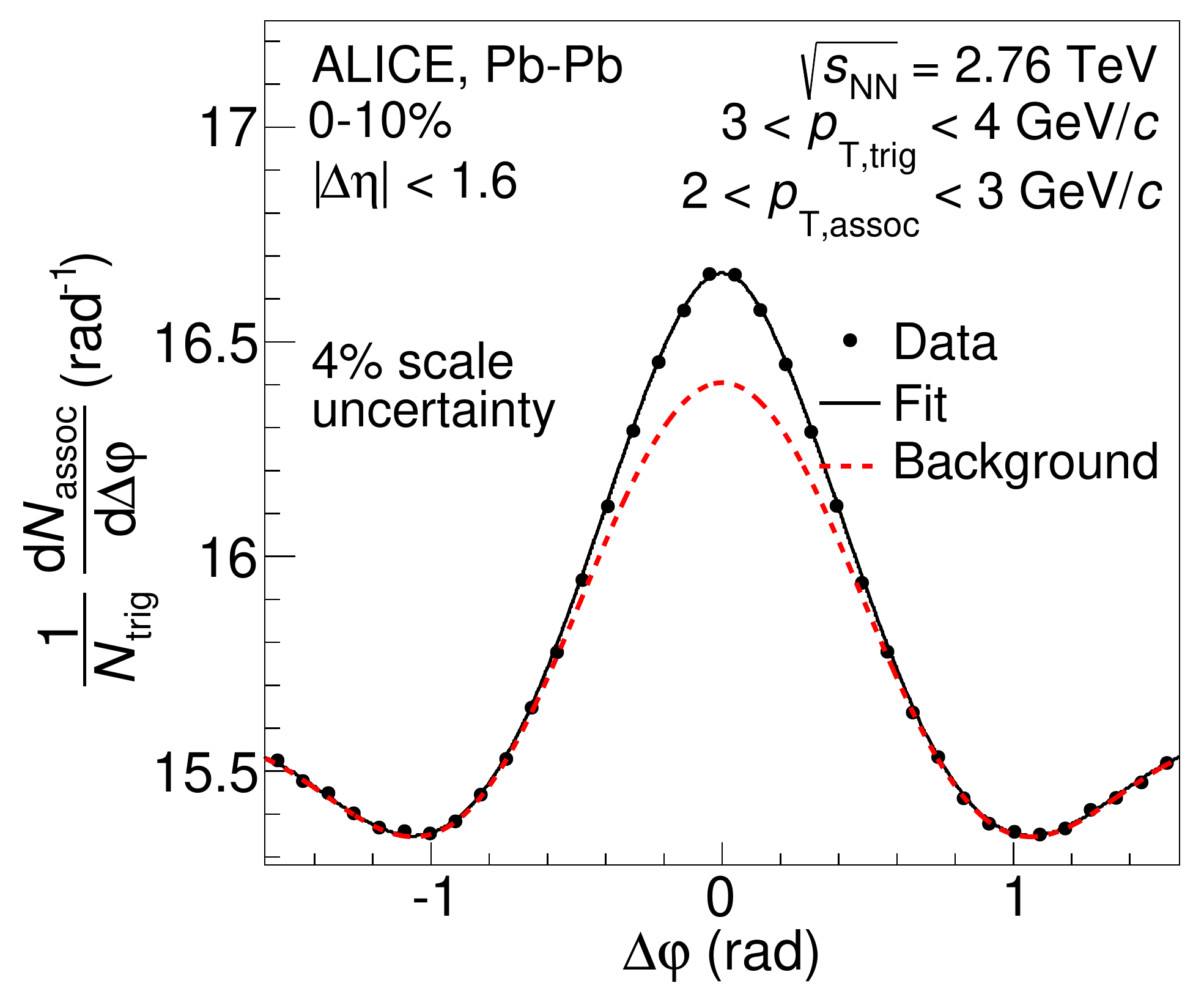}
      \label{subfig:results2c_projPhi}
    \end{overpic}}
    \subfloat[]{%
    \begin{overpic}[width=0.25\textwidth]{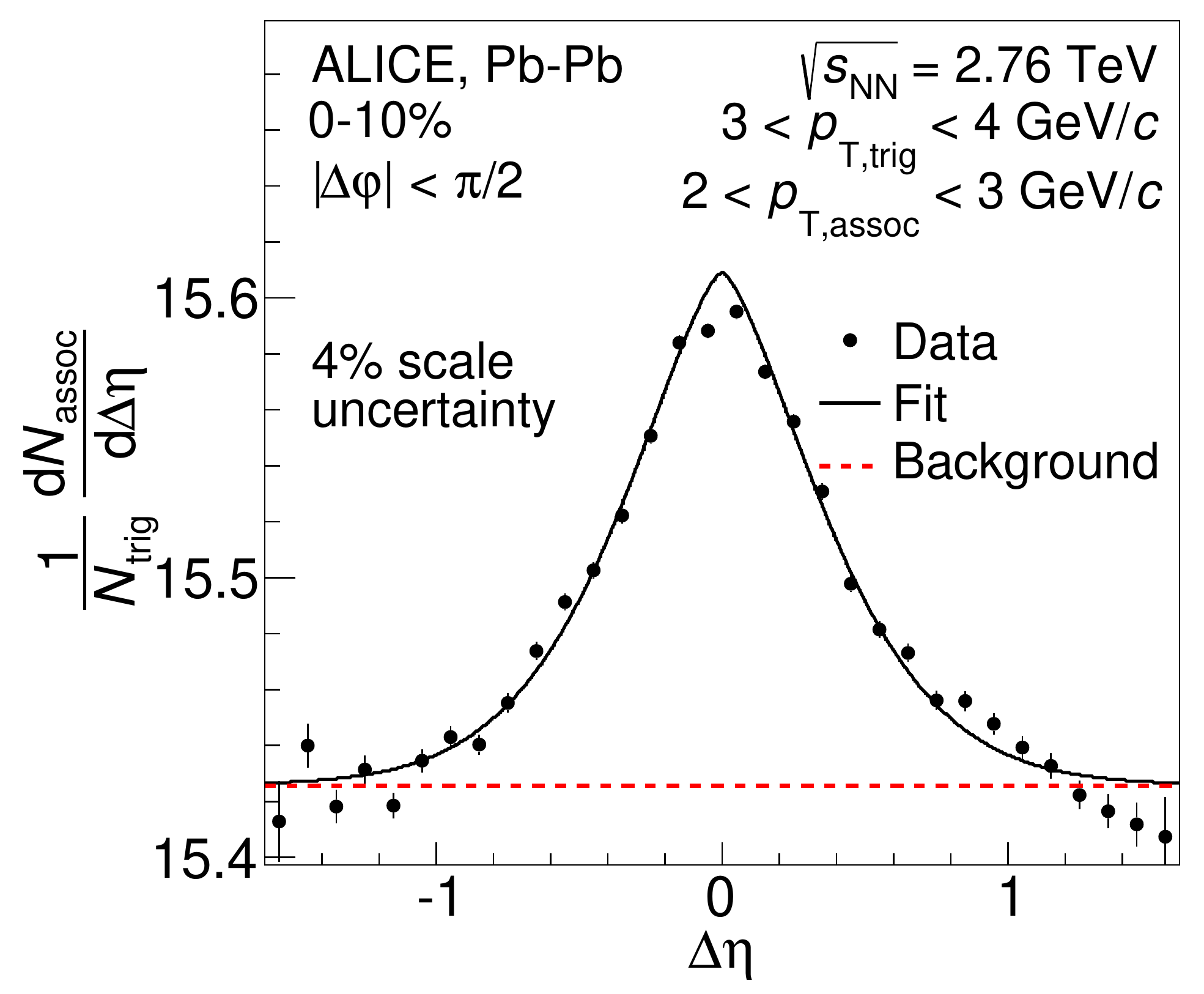}
      \label{subfig:results2c_projEta}
    \end{overpic}}}
    \caption{Projection of the per trigger yield histogram on the \Dphi \protect\subref{subfig:results2c_projPhi} and the \Deta \protect\subref{subfig:results2c_projEta} direction in Pb--Pb collisions at \snn in the most central 10\% of the events in the bin \mbox{$3<p_{\,\text{T,trig}}<4$\GeVc} and \mbox{$2<p_{\,\text{T,assoc}}<3$\GeVc}. The data is overlaid with the fitted function in black and with the background part of the fit in red.}
    \label{fig:results2c_proj}
\end{figure}

The present study focuses on the evolution of the shape of the near-side peak as a function of \pT and centrality. To provide a quantitative description, the two-dimensional per trigger yield histograms are fitted by a function which describes both the peak and the background. The background is modeled by a constant for the combinatorial part and by cosine terms up to fourth order to account for the flow modulated part \cite{Aamodt:2011by}. The peak is characterized by a generalized Gaussian in both the \Dphi and the \Deta direction:
\begin{multline}
  F(\Delta\varphi,\Delta\eta) = C_1 + \sum_{n=2}^4 2 V_{n\Delta} \cos (n \Delta\varphi) + \\
    + C_2 \cdot G_{\gamma_{\Delta\varphi},w_{\Delta\varphi}}(\Delta\varphi) \cdot G_{\gamma_{\Delta\eta},w_{\Delta\eta}}(\Delta\eta) \label{eq:fit}
\end{multline}
\begin{equation}
  G_{\gamma_x,w_x}(x) = \frac{\gamma_x}{2{w}_x\Gamma (1/\gamma_x)} \exp \left[ -\left(\frac{|x|}{w_x}\right)^{\gamma_x} \right]
\end{equation}
In Fig.~\ref{fig:results2c_proj}, the projection of an example \pT and centrality bin is shown to illustrate the fit and the magnitude of the background.

\section{Results}
\begin{figure}[!htbp]
  \begin{overpic}[width=0.51\textwidth]{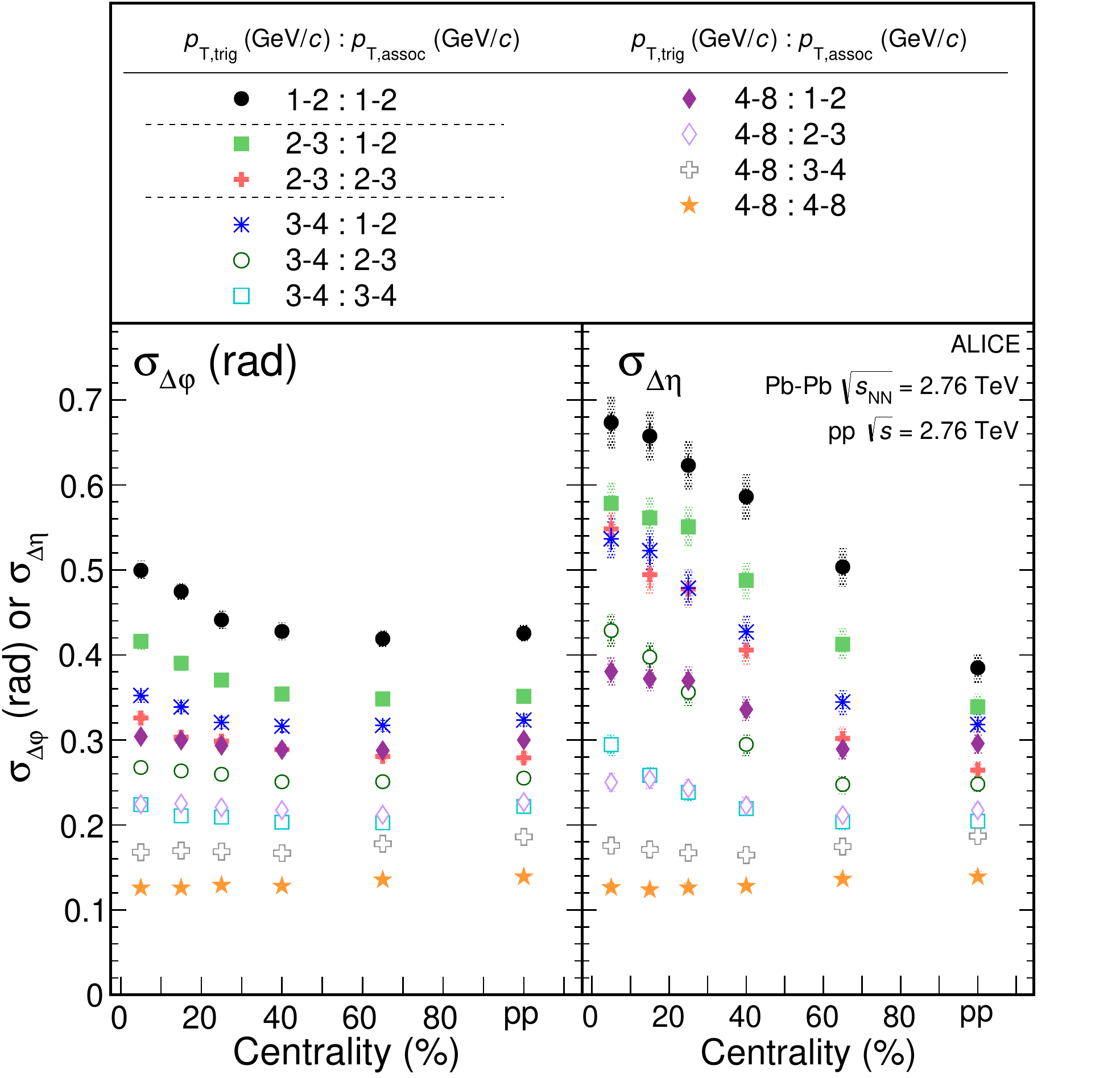}
  \end{overpic}
  \caption{Width of the near-side peak in the \Dphi (left panel) and \Deta (right panel) direction in Pb--Pb and pp collisions (rightmost points) at \snn. Lines indicated the statistical uncertainties (mostly smaller than the size of the symbols), while the filled areas represent the systematic ones.}
  \label{fig:widthTwoPanel}
\end{figure}

\begin{figure}[!htbp]
  \begin{overpic}[width=0.48\textwidth]{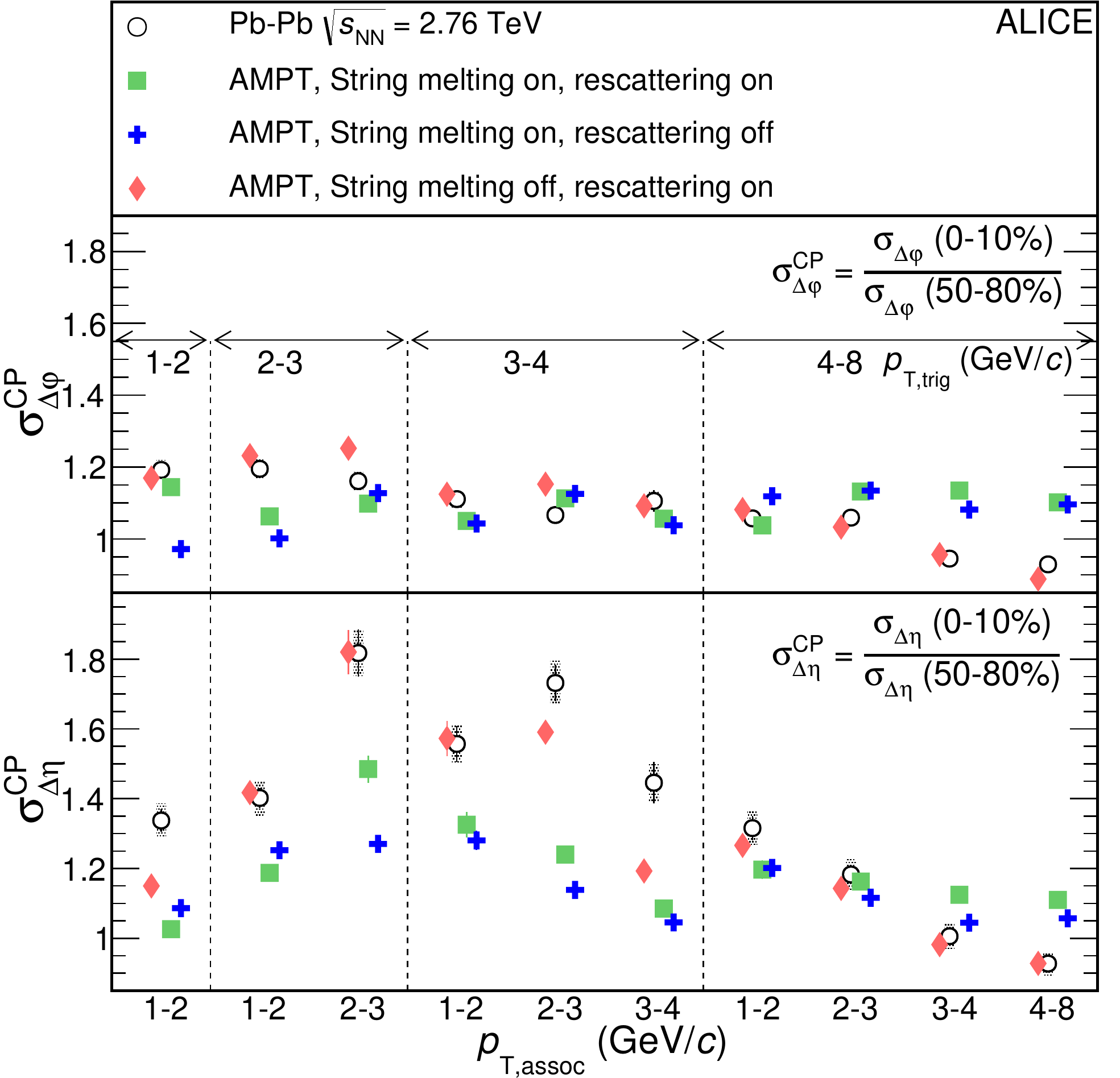}
  \end{overpic}
  \caption{Ratio of the width in the most central (0--10\%) and the most peripheral (50--80\%) bin. The x-axis contains both the \pTtrig and the \pTassoc, therefore a uniform trend is not expected. Lines indicated the statistical uncertainties (mostly smaller than the size of the symbols), while the filled areas represent the systematic ones.}
  \label{fig:CP_with2Panels}
\end{figure}

The width of the near-side peak is characterized by the variance of the fitted generalized Gaussian functions, and is presented in Fig.~\ref{fig:widthTwoPanel} as a function of \pT and centrality. Besides the ordering of the width according to \pT, a broadening towards central events is observed, which is more pronounced in the \Deta direction, and disappears above \mbox{$4<p_{\,\text{T,trig}}<8$\GeVc} and \mbox{$3<p_{\,\text{T,assoc}}<4$\GeVc}. This broadening is further studied by calculating the ratio of the width in the most central (0--10\%) bin and the most peripheral (50--80\%) bin. This $\sigma^{CP}$ is shown in Fig.~\ref{fig:CP_with2Panels} as a function of \pT. In the \Dphi direction a moderate increase of the width (at maximum 1.2) can be seen, while in the \Deta directions, the ratio goes up to around 1.8 at intermediate \pT (\mbox{$2<p_{\,\text{T,trig}}<4$\GeVc} and \mbox{$2<p_{\,\text{T,assoc}}<3$\GeVc}).

In Fig.~\ref{fig:CP_with2Panels}, the data is overlaid with results from generator level AMPT simulations~\cite{Lin:2004en, Xu:2011fi} to study the effects of the flowing medium on the shape of the peak. Three settings of AMPT have been studied, where either string melting or hadronic rescattering or both effects are activated. The version where string melting is turned off, but hadronic rescattering is active follows the data very well, while the other two settings underestimate the broadening in the \Deta direction for most \pT bins. The \Dphi direction is less constraining; however, the same AMPT productions that follows the trend in the \Deta direction, describes the results at high \pT well, while the other two settings do not. 

\begin{figure}[!htbp]
  \makebox[0.5\textwidth][c]{
    \subfloat[]{%
    \begin{overpic}[width=0.25\textwidth]{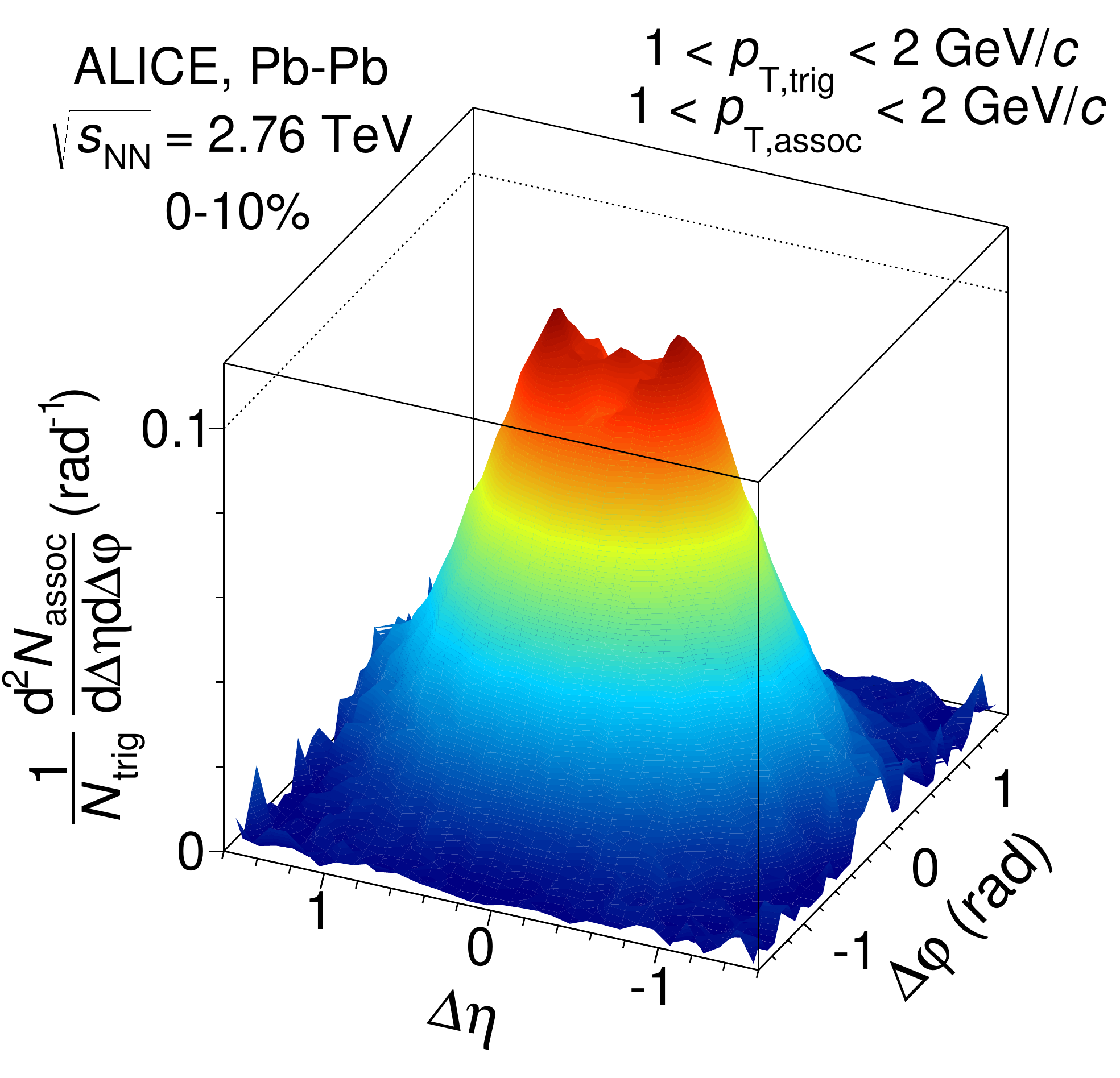}
      \label{subfig:results1c}
    \end{overpic}}
    \subfloat[]{%
    \begin{overpic}[width=0.25\textwidth]{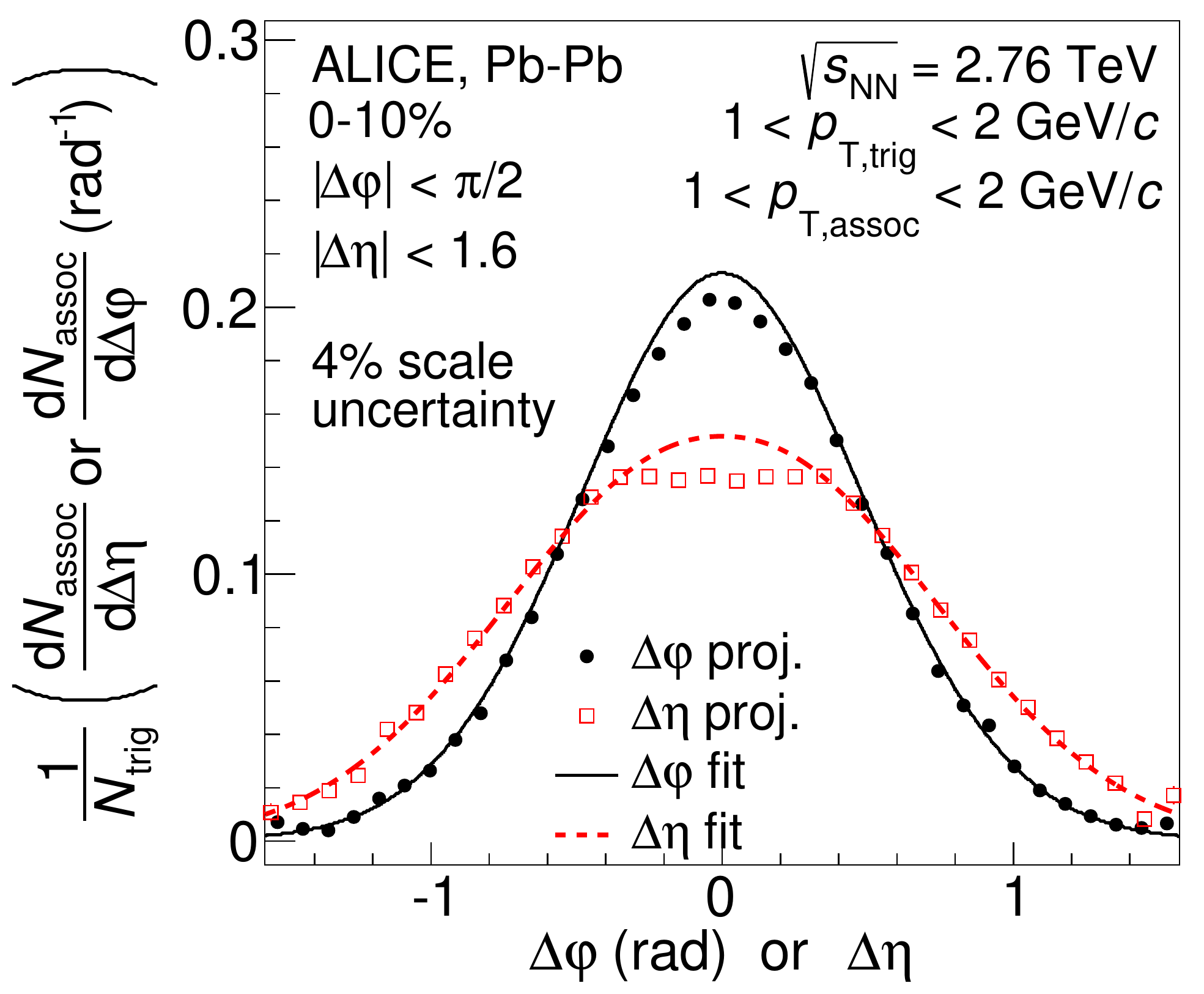}
      \label{subfig:Bothresults1c}
    \end{overpic}}}
    \caption{Panel \protect\subref{subfig:results1c} shows the per trigger yield histogram after the subtraction of the background in the lowest \pT bin (\mbox{$1<p_{\,\text{T,trig}}<2$\GeVc} and \mbox{$1<p_{\,\text{T,assoc}}<2$\GeVc}) for the 10\% most central events in Pb--Pb collisions at \snn. Panel \protect\subref{subfig:Bothresults1c} shows the projection of the same histogram to the \Dphi and \Deta direction.}
    \label{fig:results1c}
\end{figure}

\begin{figure}[!htbp]
  \begin{overpic}[width=0.48\textwidth]{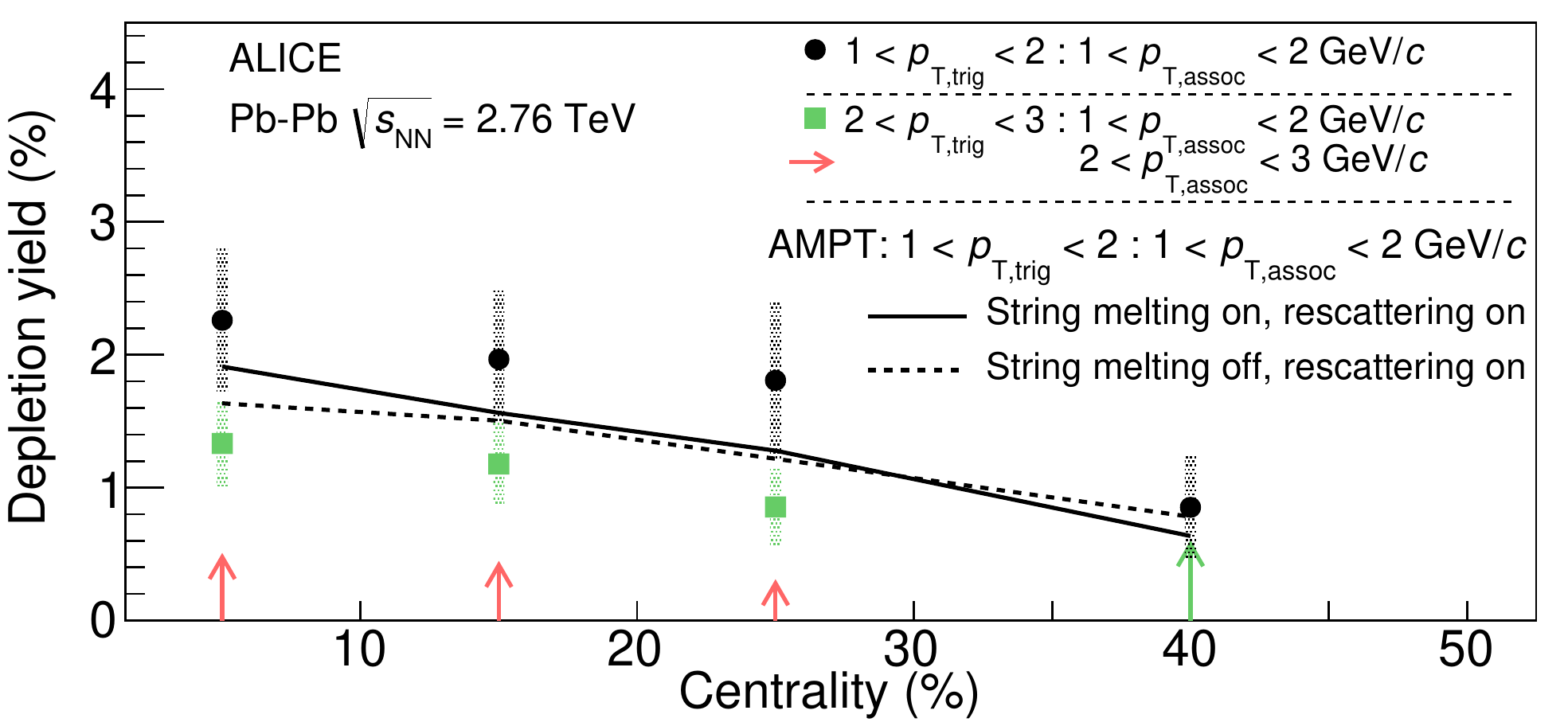}
  \end{overpic}
  \caption{Depletion yield as a function of centrality, overlaid with the AMPT cases, where it is different from 0. Lines indicated the statistical uncertainties (mostly smaller than the size of the symbols), while the filled areas represent the systematic ones. }
  \label{fig:depletion_AMPT_comparison_PRL}
\end{figure}

At low \pT in central collisions a novel feature develops, the peak departs from the Gaussian shape, and a depletion around $(\Delta\varphi,\Delta\eta) = (0,0)$ becomes visible. This is illustrated in Fig.~\ref{fig:results1c}. The area of the depletion is excluded from the previously described fit to allow for an unbiased characterization of the width. The same area can be used to characterize the depletion, which is done by taking the difference between the fit and the data and normalizing it by the total yield of the peak. This depletion yield is shown in Fig.~\ref{fig:depletion_AMPT_comparison_PRL}. It is significant in the lowest two \pT bin (\pTtrig $<3$\GeVc and \pTassoc $<2$\GeVc), and it is highest in the most central bin, where around 2.2\% of the yield is missing from the peak. No depletion is seen in the peripheral or the pp case. 

In AMPT, only those cases show a depletion where hadronic rescattering is turned on. From these two AMPT productions, the depletion yield is calculated in the same way as for data, and it is shown as lines in Fig.~\ref{fig:depletion_AMPT_comparison_PRL}. It can be seen that the depletion yield is almost independent of string melting, and that it agrees well with the data in the lowest \pT bin. In AMPT, no depletion is seen at higher \pT.

\section{Interpretation}
To study the observed effects in the context of flow, the radial expansion velocity ($\beta_{\rm T}$) and the elliptic flow coefficient ($v_2\{2\}$) are extracted from the data and the AMPT samples. For the extracted values the reader is referred to \cite{Adam:2016ckp,Adam:2016tsv}. Elliptic flow is quite well described by AMPT; however, $\beta_{\rm T}$ is underestimated by the model for all studied settings. The settings with either hadronic rescattering or string melting turned on leads to a $v_2\{2\}$ value closest to the data, with 9\% and 7\% difference respectively. Out of these two, only the setting with hadronic rescattering shows a depletion and follows the broadening trend of the data. This is also the sample, which shows the closest $\beta_{\rm T}$ value to the data with around 17\% difference. This suggests that the effects are driven rather by radial flow than by elliptic flow.


\section{Conclusions}
The paper reports about two-particle angular correlation measurements in Pb--Pb and pp collisions at \snn. It was shown that the near-side peak becomes wider towards central collisions at low \pT, and that this broadening is more pronounced in the \Deta direction than in the \Dphi direction. At low \pT and in central collisions another feature develops as well: a depletion around $(\Delta\varphi,\Delta\eta) = (0,0)$ appears, in which up to 2.2\% of the yield of the peak is missing. The data were compared to AMPT calculations with different settings, which suggested that radial flow is responsible for the observed features. In previous theoretical work~\cite{Ma:2008nd}, it was observed that large longitudinal flow in AMPT results in longitudinal broadening, therefore a possible interpretation, based on AMPT, is that the observed effects are caused by the interplay of the flowing medium with the jets.




\nocite{*}
\bibliographystyle{elsarticle-num}
\bibliography{Kofarago_M}







\end{document}